\documentclass[10pt,aps,prb,twocolumn,superscriptaddress]{revtex4-1}
\usepackage{graphicx}
\usepackage{amssymb}
\usepackage{amsmath}
\usepackage{appendix}
\usepackage{comment}

\def\p{\partial}
\def\i{\imath}
\def\j{\jmath}

\def\bc{\mathbf{c}}
\def\bm{\mathbf{m}}
\def\br{\mathbf{r}}
\def\bj{\mathbf{j}}

\def\bv{\mathbf{v}}
\def\bz{\mathbf{z}}

\def\bA{\mathbf{A}}

\def\bS{\mathbf{S}}
\def\BN{\boldsymbol{\nabla}}

\def\mA{\mathcal{A}}
\def\mG{\mathcal{G}}
\def\mE{\mathcal{E}}
\def\mH{\mathcal{H}}
\def\mL{\mathcal{L}}
\def\mO{\mathcal{O}}
\def\mR{\mathcal{R}}
\def\fa{\mathfrak{a}}
\newcommand{\rf}[1]{(\ref{#1})}
\newcommand{\al}[1]{\begin{aligned}#1\end{aligned}}

\newcommand{\eq}[1]{\begin{equation}#1\end{equation}}

\usepackage{mdframed}

\begin{document}

\title{Quantum hydrodynamics of spin winding}

\author{Yaroslav Tserkovnyak}
\author{Ji Zou}
\affiliation{Department of Physics and Astronomy, University of California, Los Angeles, California 90095, USA}
\author{Se Kwon Kim}
\affiliation{Department of Physics, Korea Advanced Institute of Science and Technology, Daejeon 34141, Republic of Korea}
\author{So Takei}
\affiliation{Department of Physics, Queens College of the City University of New York, Queens, New York 11367, USA}

\begin{abstract}
An easy-plane spin winding in a quantum spin chain can be treated as a transport quantity, which propagates along the chain but has a finite lifetime due to phase slips. In a hydrodynamic formulation for the winding dynamics, the quantum continuity equation acquires a source term due to the transverse vorticity flow. The latter reflects the phase slips and generally compromises the global conservation law. A linear-response formalism for the nonlocal winding transport then reduces to a Kubo response for the winding flow along the spin chain, in conjunction with the parasitic vorticity flow transverse to it. One-dimensional topological hydrodynamics can be recovered when the vorticity flow is asymptotically small. Starting with a microscopic spin-chain formulation, we focus on the asymptotic behavior of the winding transport based on the renormalized sine-Gordon equation, incorporating phase slips as well as Gilbert damping. A generic electrical device is proposed to manifest this physics. We thus suggest winding conductivity as a tangible concept that can characterize low-energy dynamics in a broad class of quantum magnets.
\end{abstract}

\maketitle

\section{Introduction}

In addition to efficient heat transport carried by spin dynamics along electrically-insulating spin chains,\cite{sunPRL09,*hohenseePRB14,*chernyshevPRL16,*chenAFM20} there has also been much interest in their transmission of spin signals.\cite{meierPRL03mt,*sirkerPRB11,*karraschPRB15,*langePRB18,*bulchandaniPRB18,*biellaNATC19} In the case of spin currents polarized along a direction of axial symmetry, the spin signals can propagate ballistically or diffusively, while generally also undergoing decay due to spin-nonconserving perturbations. Alternatively, transport based on collective order-parameter dynamics and rooted in topological conservation laws has been suggested for potentially more robust propagation of signals.\cite{tserkovJAP18}

The winding dynamics of planar spins in an easy-plane (anti)ferromagnet is one ready example of this. Extending the natural superfluid analogy for the SO(2) order parameter to the nonequilibrium setting, scenarios for spin superfluidity have been proposed \cite{halperinPR69,*soninJETP78,*konigPRL01,*takeiPRL14,*takeiPRL16} and experimentally pursued.\cite{yuanSA18,*stepanovNATP18} The spontaneously broken U(1) symmetry is replaced here by the axial symmetry (say along the $z$ axis) of the easy-plane spin winding (in the $xy$ plane). If the latter experiences some anisotropies within the $xy$ plane, however, the associated SO(2) symmetry gets broken directly, invalidating spin conservation and possibly pinning the conjugate phase (i.e., the winding angle) altogether.

While the spin density $\rho_z$ is then no longer acting as a long-wavelength transport quantity, the winding density $\rho\propto\p_x\varphi$ ($x$ being the spatial coordinate along the transport channel and $\varphi$ the azimuthal angle of the order parameter in the $xy$ plane) obeys a continuity equation (with the associated flux $j\propto-\p_t\varphi$), irrespective of the anisotropies. This is crucially contingent on the ability to unambiguously define $\varphi(x,t)$ along the channel, at all times, which is compromised whenever a vectorial order parameter traverses one of the poles along the hard ($z$) axis. Such processes could be visualized as vortices in the $(1+1)$-dimensional space-time, realizing a vorticity flow transverse to the $x$ axis. In analogy to similar parasitic events in low-dimensional superfluids and superconductors,\cite{halperinIJMPB10} these can be called \textit{phase slips.}\cite{kimPRB16,kimPRL16}

In this paper, we set out to formulate a rigorous microscopic formalism to address these issues, in regard to quantum winding hydrodynamics, at an arbitrary temperature. Once the formal framework is in place, our focus is going to be on the role of anisotropies, phase slips, and general magnetic damping, in relation to spatiotemporal transport properties of the spin-winding flows. In particular, we wish to establish regimes, where the notion of a \textit{winding conductivity} can be meaningful both theoretically and experimentally.

Our discussion is structured as follows. We start, in Sec.~\ref{2d}, by recapping vorticity dynamics in two spatial dimensions. The notion of a topological conservation law is introduced for a classical theory in Sec.~\ref{2dc}, which is then discretized and quantized into an exact quantum formulation on a generic spin lattice, in Sec.~\ref{2dq}. A similar procedure is then attempted for the winding dynamics in Sec.~\ref{1d}, where the quantum flow of spin winding along a spin chain gets supplemented with vorticity flow transverse to it. Here, we develop a quantum Kubo formalism for the winding transport, and establish boundary conditions that could allow us to read it out electrically. In Sec.~\ref{sgm}, a sine-Gordon model is treated systematically, in order to study the interplay of the winding flow, phase slips, and other sources of dissipation associated with collective dynamics, both at zero and finite temperatures. A summary and outlook are offered in Sec.~\ref{d}.

\section{2D vorticity (hydro)dynamics}
\label{2d}

\subsection{Classical vorticity dynamics}
\label{2dc}

A three-component real vector field $\bm=(m^x,m^y,m^z)$ residing in $2+1$ dimensions, $\bm(\br,t)$, realizes an $\mathbb{R}^2\to\mathbb{R}^3$ mapping, at any given time $t$. These spatial field textures are devoid of point defects, as the fundamental homotopy group of the order-parameter space $\bm$ is trivial: $\pi_1(\mathbb{R}^3)=1$. Such two-dimensional textures are, furthermore, all topologically equivalent, having fixed the boundary profile of $\bm$ on a connected patch of $\mathbb{R}^2$, which is reflected in the fact that $\pi_2(\mathbb{R}^3)=1$. Despite this, a smooth vector field defines a \textit{topological hydrodynamics} governed by the continuity equation $\p_\mu j^\mu=0$ (with the Einstein summation implied over the Greek letters: $\mu=0,1,2\to t,x,y$), where\cite{zouPRB19}
\eq{
j^\mu\equiv\frac{\epsilon^{\mu\nu\xi}\,\bz\cdot\p_\nu\bm\times\p_\xi\bm}{2\pi}\,.
\label{jmu}}
Here, $\bz$ is the z-axis unit vector and $\epsilon^{\mu\nu\xi}$ is the Levi-Civita symbol.

For the special case of a rigid texture sliding at a velocity $\bv$, for example: $\bj=\rho\bv$, where $\rho\equiv j^0$ and $\bj=(j^x,j^y)$. For another special case of a sharp vortex in a strongly easy-plane magnet with the planar order parameter normalized to unity, $|\bm|\to1$: $\rho\approx\delta(\br-\br_0)$, where $\br_0$ is the position at which $\bm$ tilts out of the plane (over an appropriate healing length defining the size of the core). These examples intuitively suggest a fluid whose density is given by the distribution of vorticity in the system. While in the extreme easy-plane case, a vortex core carries a quantized topological charge, we do not generally assume this special limit.

The above conserved quantity $j^0$ can be recast as a fictitious flux
\eq{
\rho=\frac{\bz\cdot\BN\times\bA}{2\pi}
\label{rho}}
associated with the \textit{gauge field}
\eq{
\bA=m^x\BN m^y-m^y\BN m^x\,.
\label{A}}
Applying Green's theorem, we then see that the conserved \textit{topological charge} within a patch $\Omega$,
\eq{
Q\equiv\int_\Omega d^2r\,\rho=\oint_{\p\Omega}\frac{d\br\cdot\bA}{2\pi}=\oint_{\p\Omega}\frac{d\phi}{2\pi}\,\bm^2_\parallel\,,
\label{Q}}
is associated with the order-parameter winding around its boundary $\p\Omega$. $\bm_\parallel$ is the field's projection onto the $xy$ plane (within the order-parameter space) and $\phi$ is the associated azimuthal angle. This reveals the geometrical meaning of the conservation law: The charge $Q$ in the bulk can change only in response to a vorticity flow through the boundary.

\subsection{Quantum vorticity dynamics}
\label{2dq}

\begin{figure}[!th]
\includegraphics[width=0.7\linewidth]{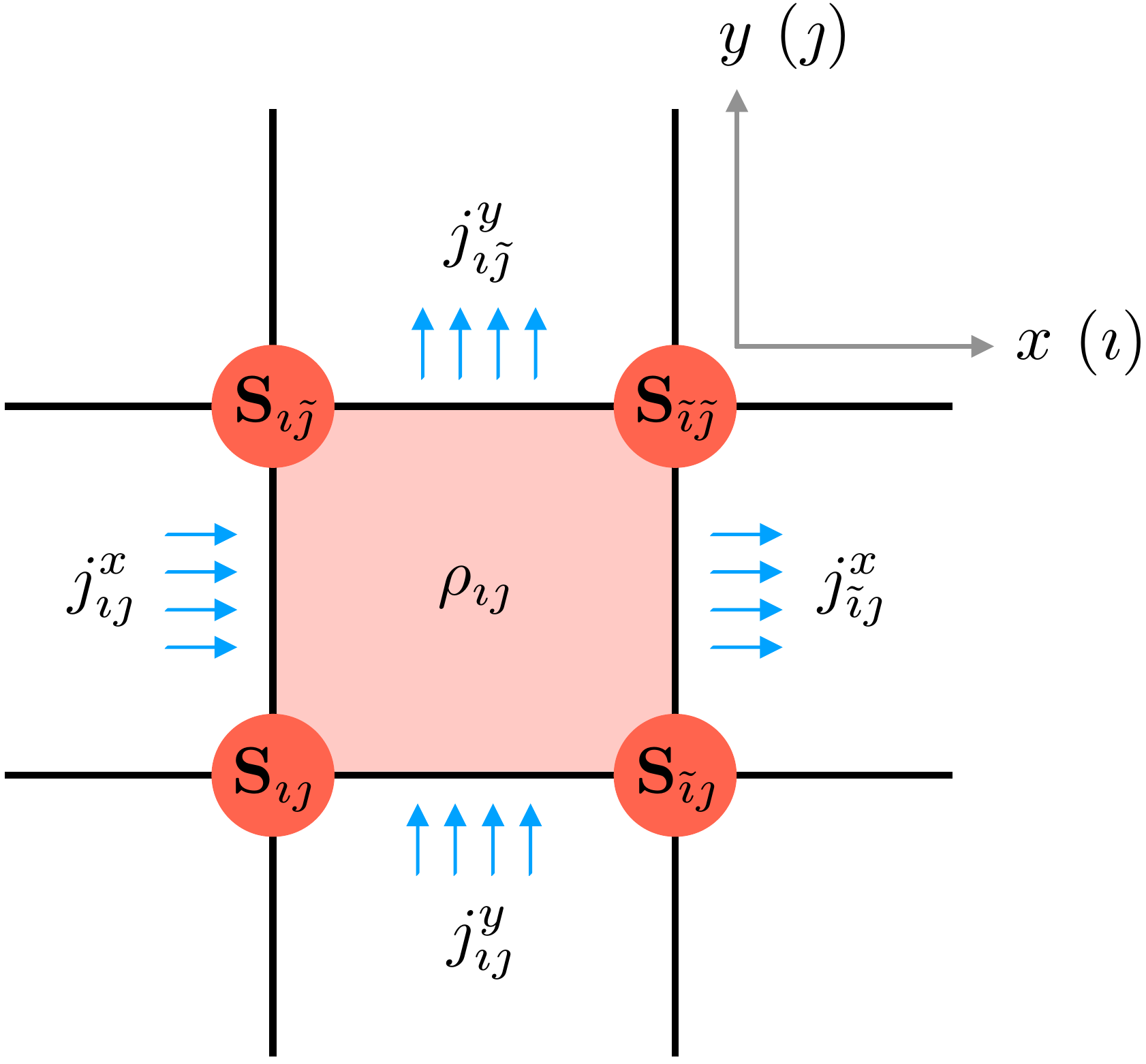}
\caption{The quantum spin lattice described by an arbitrary Hamiltonian $H$. $\bS_{\i\j}$ is the spin operator at site $\i\j$, with index $\i$ ($\j$) running along the $x$ ($y$) axis. $\tilde{\i}=\i+1$ and $\tilde{\j}=\j+1$. $\rho_{\i\j}$ is the conserved topological charge per plaquette $\i\j$, $j^x_{\i\j}$ ($j^y_{\i\j}$) is the flux per vertical (horizontal) link $\i\j$, which together satisfy the quantum continuity equation \rf{CE}.}
\label{sch}
\end{figure}

To construct a simple quantum theory, which reproduces the above classical hydrodynamics of vorticity in the classical limit of $\hbar\to0$, let us consider a square lattice model sketched in Fig.~\ref{sch}. We label each vertex of the lattice by two integer indices: $\i$ (along the $x$ axis) and $\j$ (along the $y$ axis). The same indices are used to label the square plaquettes, according to their lower left corner, as well as the vertical links going upward and the horizontal links to the right of the site $\i\j$. Each site contains a quantum spin $\bS=(S^x,S^y,S^z)$, of magnitude $S$ (in units of $\hbar$), characterized by the standard angular-momentum algebra $[S^a,S^b]=i\epsilon^{abc}S^c$.

We associate a charge density
\eq{
\rho_{\i\j}\equiv\frac{A^x_{\i\j}-A^x_{\i\tilde{\j}}+A^y_{\tilde{\i}\j}-A^y_{\i\j}}{2\pi a}
\label{rhoijA}}
to each plaquette, where $a$ is the lattice spacing. Here, $\tilde{\i}\equiv\i+1$ and $\tilde{\j}\equiv\j+1$, and
\eq{\al{
A^x_{\i\j}&=\frac{\bz\cdot(\bS_{\tilde{\i}\j}+\bS_{\i\j})\times(\bS_{\tilde{\i}\j}-\bS_{\i\j})}{4aS^2}+{\rm H.c.}=\frac{\bz\cdot\bS_{\i\j}\times\bS_{\tilde{\i}\j}}{aS^2}\,,\\
A^y_{\i\j}&=\frac{\bz\cdot(\bS_{\i\tilde{\j}}+\bS_{\i\j})\times(\bS_{\i\tilde{\j}}-\bS_{\i\j})}{4aS^2}+{\rm H.c.}=\frac{\bz\cdot\bS_{\i\j}\times\bS_{\i\tilde{\j}}}{aS^2}\,,
\label{QA}}}
which we assign formally to the corresponding horizontal and vertical sides of the plaquette, respectively. These definitions mimic Eqs.~\eqref{rho} and \eqref{A}, respectively, and should reproduce them by coarse graining the magnetic textures in the classical limit of $S\to\infty$.

According to these definitions,
\eq{
\rho_{\imath\jmath}=\frac{\bz\cdot\bc_{\imath\jmath}}{2\pi a^2}\,,~~~{\rm where}~~~\bc_{\imath\jmath}\equiv\frac{1}{S^2}\sum_l\bS_l\times\bS_{\tilde{l}}
\label{rhoij}}
is the \textit{vector chirality} of the corresponding plaquette, with the sum running over the four vertices labelled by $l$ ($\tilde{l}$ being the vertex next to $l$, in the counterclockwise direction).\cite{Note1} We also see [according to Eq.~\rf{rhoijA}] that
\eq{
Q=\sum_{\i\j}\rho_{\i\j}
}
vanishes in the bulk and reduces to the boundary terms, which we can interpret as the quantum version of the vorticity \eqref{Q}. This suggests a conservation law with the boundary fluxes corresponding to the vorticity flow. Indeed, according to the Heisenberg equation of motion (for Hamiltonian $H$ and an arbitrary time-independent operator $\mO$),
\eq{
\p_t\mO\equiv\frac{i}{\hbar}[H,\mO]\,,
\label{EOM}}
the quantum vorticity density $\rho_{\i\j}$ is seen to satisfy the continuity equation:
\eq{
\p_t\rho_{\i\j}+\frac{j^x_{\tilde{\i}\j}-j^x_{\i\j}+j^y_{\i\tilde{\j}}-j^y_{\i\j}}{a}=0\,.
\label{CE}}
The fluxes in the second term are consistent with quantizing Eq.~\rf{jmu}:
\eq{
j^x_{\i\j}=\frac{\bz\cdot(\bS_{\i\tilde{\j}}-\bS_{\i\j})\times\p_t(\bS_{\i\tilde{\j}}+\bS_{\i\j})}{4\pi aS^2}+{\rm H.c.}\,,
\label{jij}}
and similarly for the other components.

It is useful to emphasize that the associated conservation law is not rooted in any specific symmetry of the system. Indeed, the form of the Hamiltonian $H$ still remains arbitrary. The continuity is rather dictated by the topology associated with the vorticity (hydro)dynamics in the interior of the system. Specifically, an arbitrary local deformation of the field in the bulk yields the same net vorticity, irrespective of the details of the dynamics.

\section{1D Winding dynamics}
\label{1d}

In contrast to the vorticity flow, winding dynamics in, e.g., one-dimensional (1D) superfluids\cite{halperinIJMPB10} or magnets\cite{kimPRB15br} obey the conservation law only approximately. In these systems, the underlying topological invariant relies on a nonlinear constraint applied to the order parameter, which ultimately makes the conserved quantity vulnerable to thermal fluctuations. This leads to phase slips,\cite{kimPRB16} which are detrimental to the topological conservation law.

These issues carry over to the quantum regime, where quantum phase slips arise due to tunneling.\cite{kimPRL16} Supposing these could be neglected, in an appropriate limit, we wish to formulate a Kubo approach for the topological quantum flow in terms of the corresponding current autocorrelator.

\subsection{Quantum winding dynamics}
\label{1da}

Let us illustrate these points by considering winding dynamics along a 1D quantum lattice, with an easy-plane anisotropy in spin space, which constrains the (ferro- or antiferro-)magnetic dynamics to lie close to the $xy$ plane. As we have already mentioned, a coarse-grained classical hydrodynamics can be formulated in terms of the density $\rho=-\p_x\varphi/\pi$ and flux $j=\p_t\varphi/\pi$, where $\varphi$ is the azimuthal angle of the order parameter in the $xy$ plane,\cite{kimPRB15br} such that $\p_t\rho+\p_xj=0$.

Allowing for arbitrary (unconstrained) spin dynamics, we now formulate a quantum theory on a lattice through the definitions
\eq{
\rho_{\i}=\frac{\bz\cdot\bS_{\tilde{\i}}\times\bS_{\i}}{\pi aS^2}\,,\,\,\,j_{\i}=\frac{\bz\cdot\bS_{\i}\times\p_t\bS_{\i}}{2\pi S^2}+{\rm H.c.}\,,
\label{rhoj}}
where, as before, $\p_t$ should be understood according to the Heisenberg equation of motion \rf{EOM} (which depends on a concrete Hamiltonian, to be specified later). Since these reduce to the winding density and flux, in the appropriate coarse-grained classical limit, we may expect them to approximately obey the continuity equation (when the phase slips can be disregarded). Indeed,
\eq{
\p_t\rho_{\i}+\frac{j_{\tilde{\i}}-j_{\i}}{a}=\frac{\bz\cdot(\bS_{\tilde{\i}}-\bS_{\i})\times\p_t(\bS_{\tilde{\i}}+\bS_{\i})}{2\pi aS^2}+{\rm H.c.}\,.
\label{CEw}}
The term on the right-hand side (RHS), which spoils the exact conservation law, can be recognized to be exactly (twice) the vorticity flow transverse to the spin chain, cf. Eq.~\rf{jij}. If it can be neglected, we would recover the continuity equation and with it the Kubo formula \rf{jj} that governs the topological flow and the electrical transconductance, to be discussed below.

\subsection{Boundary conditions}
\label{1db}

In order to place the spin chain (of length $L$) into a measurable external circuit, let us suppose it is biased by spin torques (polarized along the $z$ axis) $\tau_{L(R)}$ at its left (right) ends. The (semiclassical) work associated with the left torque is
\eq{
\Delta W_L=\int dt \tau_L\partial_t\varphi=\pi\tau_L\int dt j=\pi\tau_L\Delta Q_L\,,
}
where $\Delta Q_L$ is the topological charge transfer into the chain through the left end. This translates into the effective chemical potential bias at the left end given by
\eq{
\mu_L=\frac{\Delta W_L}{\Delta Q_L}=\pi\tau_L\,.
}
Similarly for the right end, we get
\eq{
\mu_R=\frac{\Delta W_R}{\Delta Q_R}=-\pi\tau_R\,.
}
Such torques can be induced, for example, by the spin Hall effect triggered by an electrical current flowing transverse to the chain.\cite{takeiPRL14} See Fig.~\ref{circ} below.

Reciprocally to these torques, the precessional dynamics produces a transverse motive force on the electrons in the contacts, which can be used to detect the outflow of the topological charge through the ends.\cite{takeiPRL14,kimPRB15br} We will return to discuss this in more detail in Sec.~\ref{1dd}.

\subsection{Kubo formula}
\label{1dc}

We are now ready to define the bulk impedance for the topological flow, as an intrinsic property of the quantum magnet. Starting with a continuity equation for the coarse-grained quantum dynamics in the bulk, we have
\eq{
\p_t\rho+\p_x j=0\,,
\label{wce}}
where the conserved density and current are obtained from Eqs.~\rf{rhoj}, and we neglect the RHS of Eq.~\rf{CEw}, i.e., phase slips, for now. We recall that the time derivatives are obtained in the Heisenberg picture. If we perturb the system by a scalar potential $\phi(x,t)$ that couples linearly to the topological charge, the Hamiltonian becomes
\eq{
H\to H+\int dx\,\phi(x,t)\rho(x)\,.
}
Note that the topological density is even under time reversal, while the flux is odd (supposing the Hamiltonian is time-reversal invariant), so it vanishes in equilibrium, when $\phi\equiv0$. For a finite time-dependent potential $\phi$, on the other hand, the linear response dictates
\eq{
j(x,t)=\frac{1}{\hbar}\int dx'dt'\mG(x-x',t-t')\phi(x',t')\,,
}
where
\eq{
\mG(x-x',t-t')\equiv-i\theta(t-t')[j(x,t),\rho(x',t')]\,,
}
according to the Kubo formula (with the equilibrium expectation value implicit on the right-hand side).

To invoke the continuity equation, we differentiate the response function in time:
\begin{widetext}
\eq{\al{
\p_t\mG(x-x',t-t')&=i\theta(t-t')[j(x,t),\p_{t'}\rho(x',t')]-i\delta(t-t')[j(x),\rho(x')]\\
&=-i\theta(t-t')[j(x,t),\p_x'j(x',t')]+\delta(t-t')\p_x'p(x-x')\,,
}}
\end{widetext}
where the auxiliary function $p(x-x')$ is obtained by integrating
\eq{
\p_x'p(x-x')=-i[j(x),\rho(x')]\,.
\label{dp}}
Fourier transforming in time, $j(\omega)=\int dte^{i\omega t}j(t)$ etc., we finally get
\eq{
j(x,\omega)=\frac{i}{\hbar\omega}\int dx'\varsigma(x-x',\omega)\varepsilon(x',\omega)\,,
}
where
\eq{\al{
\varsigma(x-x',t-t')\equiv&-i\theta(t-t')[j(x,t),j(x',t')]\\
&+\delta(t-t')p(x-x')
}\label{vs}}
involves the current autocorrelator and
\eq{
\varepsilon\equiv-\p_x\phi
}
is the effective electric field. This gives for the conductivity relating $j(k,\omega)$ to $\varepsilon(k,\omega)$:
\eq{
\sigma(k,\omega)=\frac{i}{\hbar\omega}\varsigma(k,\omega)\,,
\label{jj}}
having also Fourier transformed in real space, $\int dx\,e^{-ikx}$.

For a torque-biased spin chain,
\eq{
\varepsilon=\frac{\mu_L-\mu_R}{L}=\pi\frac{\tau_L+\tau_R}{L}\,,
}
supposing that the length of the topological transport channel $L$ is long enough, so that the bulk dominates over the interfacial impedances and focusing on the DC limit.\cite{tserkovJAP18}

While evaluating the Kubo formula, we should in general also calculate the phase-slip rate governed by the RHS of Eq.~\rf{CEw} (driven by the potential $\phi$ gradient that couples to the winding density $\rho$). For the internal consistency of the hydrodynamic treatment, it must be small compared to the induced winding flux along the spin chain. Writing the phase-slip rate per unit length (i.e., the vorticity flux transverse to the chain) as
\eq{
j_\phi=\kappa\varepsilon\,,
\label{jke}}
where $\varepsilon=-\p_x\phi$ is the effective field that drives the winding flow, we thus require (taking the $k\to0$ and $\omega\to0$ limit for $\sigma$)
\eq{
L\ll\frac{\sigma}{\kappa}\,,
\label{L}}
for the validity of the (approximate) conservation law \rf{CEw}. At the same time, however, we should not forget that $L$ must be long enough for us not to concern with the effective interfacial resistance, if we want the overall impedance to be governed by the bulk.

\subsection{Electrical transconductance}
\label{1dd}

\begin{figure}[!th]
\includegraphics[width=\linewidth]{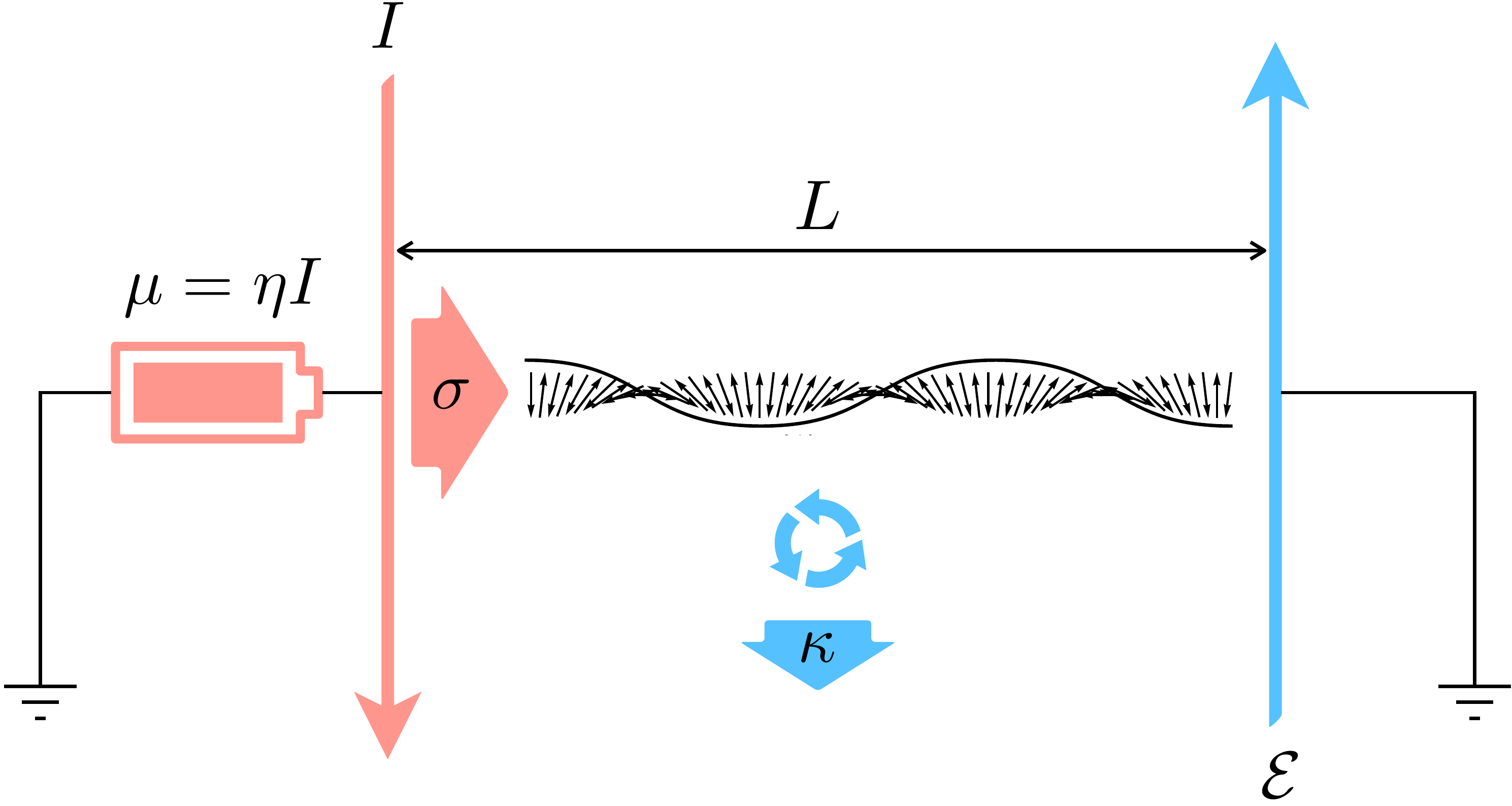}
\caption{Schematic of a winding flow along a (horizontal) spin chain. Transverse charge current $I$ generates an effective chemical potential bias $\mu$ that couples to the winding density at the left end. $\eta$ is a contact-dependent conversion parameter, which relates the input current to the out-of-plane spin Hall torque $\tau=\eta I/\pi$ acting on the magnetic dynamics in the chain.\cite{takeiPRL14} The injected winding flux is governed by the winding conductivity $\sigma$, while being dissipated by the transverse vortex flow $\propto\kappa$. The net remaining winding outflow produces a measurable transverse motive force $\mE$ at the right electrical contact.}
\label{circ}
\end{figure}

If the winding injection is performed electrically, so that the effective chemical potential (conjugate to the topological charge) $\mu=\eta I$, where $I$ is the applied current, the Onsager reciprocity dictates the backaction motive force on the electrons $\mE=\eta j$ (which translates directly into a measurable voltage).\cite{tserkovJAP18} Putting this together, for a circuit sketched in Fig.~\ref{circ}, we obtain the electrical transconductance
\eq{
G=\frac{\mE}{I}=\eta^2g\,.
\label{G}}
mediated by the winding flow across the chain. The (linear response) effective winding conductance here,
\eq{
g\equiv\frac{j}{\mu}\,,
}
would be given by $\sigma/L$ in the absence of phase slips, but is degraded by them otherwise, as a function of $L$.\cite{Note2}

\section{Sine-Gordon model}
\label{sgm}

As a concrete illustration of this general field-theoretic formalism, let us now consider an ideal 1D spin chain with an easy-plane collinear order parameter parametrized by azimuthal angle $\varphi(x)$ and the (canonically-conjugate) out-of-plane spin density $s(x)$. Our main focus is on the antiferromagnetic case (which will affect the phase-slip considerations). The classical low-energy dynamics are generated by the (sine-Gordon) Hamiltonian density
\eq{
\mH=\frac{s^2}{2\chi}+\frac{A(\partial_x\varphi)^2}{2}+K\cos(2\varphi)\,,
\label{Hs}}
along with the Poisson bracket $\{\varphi(x),s(x')\}=\delta(x-x')$. It is assumed here that the order-parameter configuration is close to the easy plane, at all times. $\chi$, $A$, and $K$ are respectively the out-of-plane spin susceptibility, order-parameter stiffness, and the axial in-plane anisotropy. The Hamiltonian is invariant under time reversal, when $s\to-s$ and $\varphi\to\varphi+\pi$. The sign of $K$ is inconsequential, as it can be flipped by a phase change, $\varphi\to\varphi+\pi/2$.

\subsection{Luttinger-liquid mapping}
\label{llba}

This description can be quantized by promoting the Poisson bracket to the commutator:
\eq{
[\varphi(x),s(x')]=i\hbar\delta(x-x')\,,
\label{fs}}
making the theory formally analogous to a spinless Luttinger liquid.\cite{giamarchiBOOK04} $s(x)$ would then correspond to the linear-momentum density $\Pi$ and the topological density to the particle density $\partial_x\phi/\pi$. Indeed, the (spinless) Luttinger-liquid Hamiltonian density is
\eq{
\mH=\frac{u}{2}\left[\frac{\pi g}{\hbar}\Pi^2+\frac{\hbar}{\pi g}(\partial_x\phi)^2\right]+K\cos(4\phi)\,,
}
where $[\phi(x),\Pi(x')]=i\hbar\delta(x-x')$. $u$ is the speed of sound and $g>0$ is the interaction parameter ($u\to v_F$, the Fermi velocity, and $g\to1$, for free electrons; $g<1$ signals electron repulsion and $g>1$ attraction). $K$ parametrizes umklapp scattering (which requires an appropriate lattice filling factor).

The corresponding Euclidean Lagrangian density is
\eq{
\mL=\frac{\hbar}{2\pi g}\left[\frac{1}{u}(\partial_\tau\varphi)^2+u(\partial_x\varphi)^2\right]+\frac{k}{2(\pi\fa)^2}\cos(2\varphi)\,,
\label{EL}}
$\fa$ being a short-distance cutoff. We have redefined the displacement field $\phi\to\varphi/2$ and appropriately rescaled $g$, in order to match the notation of our spin model \rf{Hs}. Under the Wilsonian renormalization-group (RG) rescaling,\cite{giamarchiBOOK04} we get the Kosterlitz-Thouless flow equations:
\eq{
\frac{dy}{dl}=(2-g)y\,,\,\,\,
\frac{dg}{dl}=-g^3y^2/8\,,
\label{dyg}}
where $y\equiv k/\pi\hbar u$ (which we can take to be positive, without loss of generality), and we have omitted the nonuniversal cutoff-dependent numerical prefactor on the right-hand side of the second equation. The RG flow of $y$ corresponds simply to the scaling dimension of the cosine operator.\cite{giamarchiBOOK04} The generic reduction in $g$, under the RG flow \rf{dyg}, corresponds to the stiffening of the order parameter $\varphi$ due to the cosine term $\propto K$ in the Lagrangian \rf{EL} (which tries to order the field $\varphi$).

For our original spin system, Eq.~\rf{Hs}, $u=\sqrt{A/\chi}$, $g=\hbar/\pi\sqrt{A\chi}$. We interpret the cutoff as $\fa\sim\sqrt{A/K_\perp}$,\cite{Note3} where $K_\perp\gg K$ is a strong easy-plane anisotropy that keeps spin dynamics close to the $xy$ plane. The bare order-parameter stiffness is $A\approx S^2Ja$ and spin susceptibility $\chi\approx\hbar^2/4Ja$, in the large-spin Heisenberg limit (which acquire some corrections due to quantum fluctuations when $S\sim1$). These estimates boil down to: $u\approx Ja/\hbar$, $g\approx2/\pi S$, $y\sim K/K_\perp\ll1$. Going beyond the Heisenberg limit, with a large easy-plane anisotropy, would decrease $\chi$, while not similarly affecting $A$, and thus increase the value of $g$.

Expanding $g$ close to its critical value $g_c=2$, $g\to2+g$, we get\cite{giamarchiBOOK04}
\eq{
\frac{dy}{dl}=-gy\,,\,\,\,
\frac{dg}{dl}=-y^2\,,
}
which parametrize hyperbolic trajectories with $g^2-y^2={\rm const}$. These equations flow rapidly to a noninteracting (gapless) fixed point $g^*>0$ and $y=0$, if $0<|y|<g$. Otherwise, the flow takes us to a gapped strong coupling, with $|y|\to\infty$, where the phase is pinned along the easy axis (i.e., $\varphi\to0$ or $\pi$). In this latter case, the elementary dynamics could be constructed out of the corresponding domain walls or their composites.\cite{giamarchiBOOK04}

\subsection{Quantum phase slips}
\label{llb}

It is important to also make an internal consistency check to ensure that the original spin system can indeed be described effectively by the sine-Gordon Lagrangian \rf{EL} for nonsingular fields $\varphi$ (i.e., configurations free of vortex singularities in the $1+1$ space-time dimensions). In the absence of the anisotropy $K$ (or when it renormalizes down to zero due to quantum fluctuations), the easy-plane antiferromagnetic spin chain undergoes a (Kosterlitz-Thouless type) spin superfluid-insulator transition at $S\approx2$ ($1/2$), for integer (half-integer) spin $S$,\cite{kimPRL16} with larger $S$ placing us in the (gapless) superfluid phase. In the latter, the magnetic vortices and antivortices bind, and we restore a hydrodynamic easy-plane $\sigma$ model.

The rate of quantum phase slips, $\kappa$, in this limit is given by\cite{kimPRL16}
\eq{
\kappa\propto\rho^{\pi q^2S-3}\,,
\label{kap}}
at zero temperature (with a similar scaling with temperature, at zero winding density $\rho$), where the elementary vorticity is $q=1$ (2) for the integer (half-integer) spin $S$.\cite{Note4} The suppression of the phase slips as $\rho\to0$, at low temperatures, gives us an algebraically-large window, Eq.~\rf{L}, for the spin-chain length $L$ with conserved winding transport.\cite{Note5}

Summarizing the requirements for the low-energy hydrodynamic regime described by the effective long-wavelength Hamiltonian
\eq{
\mH=\frac{s^2}{2\chi}+\frac{A(\partial_x\varphi)^2}{2}\,,
\label{Hss}}
with canonically conjugate $s$ and $\varphi$ (and effective, renormalized $\chi$ and $A$), we need $S\gtrsim1/2$ (for the half-integer case, where the leading-order phase slips are mediated by double vortices, due to destructive interference of single vortices,\cite{kimPRL16}) while $g\sim2/\pi S>2$, and $y=k/\pi\hbar u<g-2$.

The generic mean-field parameters for a Heisenberg antiferromagnetic chain\cite{Note6} thus place us at a cusp of the gapless (hydrodynamic) phase for $S=1/2$ (with a clear tendency to flow to the gapped phase for a larger spin). $g$ can be decreased beyond the Heisenberg limit when the easy-plane anisotropy is large (i.e., comparable to the exchange interaction), which would suppress the spin susceptibility $\chi$. A more detailed and careful analysis would be needed, in principle, in order to establish the exact values of the parameters yielding the ballistic transport regime in this quantum limit of $S\approx1/2$. A fermionic treatment in the Wigner-Jordan representation of spin dynamics\cite{hoffmanCM18} is one possible approach to that end.

\subsection{Winding hydrodynamics}
\label{llc}

Supposing we wind up with a gapless regime described by the Hamiltonian \rf{Hss} (if the anisotropy $K$ is weak and/or irrelevant), we are ready to formulate a transport theory for the winding density $\rho=-\partial_x\varphi/\pi$. The associated current operator, $j=\partial_t\varphi/\pi$, becomes [cf. Eq.~\rf{fs}]
\eq{
j=\frac{i}{\pi\hbar}[H,\varphi]=\frac{s}{\pi\chi}\,.
\label{js}}
From Eq.~\rf{dp}, we thus get
\eq{
\p_x'p(x-x')=-i[j(x),\rho(x')]=-\frac{\hbar}{\pi^2\chi}\p_x'\delta(x-x')\,,
}
so that
\eq{
p(x-x')=\frac{\hbar}{\pi^2\chi}\delta(x-x')\,.
}
The current-current correlator in Eq.~\rf{vs} vanishes in the limit $k\to0$ (while maintaining a finite $\omega$),\cite{tserkovPRR19} so that we end up with the conductivity \rf{jj}:
\eq{
\sigma(k\to0,\omega)=\frac{i}{\pi^2\chi\omega}=\frac{i\mA}{\omega}\,,
}
where $\mA=1/\pi^2\chi$.

Regularizing this result at zero frequency, $\omega\to\omega+i0^+$, we get ${\rm Re}\sigma=\pi\mA\delta(\omega)$. As expected, the static conductivity diverges in the low-frequency limit. In this case, the superfluid bulk has no impedance and the winding conductance of the entire structure needs to be determined by carefully considering the interfacial injection physics, which is akin to the Andreev conductance of normal$|$superconducting interfaces.\cite{nazarovBOOK09} The resultant winding conductance of a clean finite-length spin chain should thus be governed by the contact impedance, similarly to the charge conductance of a ballistic electronic wire (whose contact resistance depends both on the details of the contact itself as well as the electron-electron interactions in the wire.\cite{maslovPRB95})

If, on the other hand, we add some spin-relaxing impurities to the Hamiltonian, so that the collective spin density defining the topological current according to Eq.~\rf{js} is now allowed to relax, the dynamic spin susceptibility can be obtained
at $k\to0$ according to the Bloch phenomenology:
\eq{
\frac{ds(t)}{dt}=-\frac{s(t)-s_0(t)}{\tau}\,.
}
Here $s_0(t)=\chi h(t)$ is the (instantaneous) equilibrium spin density, according to the Zeeman coupling to a magnetic field: $\mH\to \mH-hs$. The dynamic spin susceptibility is thus:
\eq{
\chi_s(\omega)\equiv\frac{s(\omega)}{h(\omega)}=\frac{\chi}{1-i\omega\tau}\approx\chi(1+i\omega\tau)\,,
}
at $\omega\ll1/\tau$. This, in turn gives us the current \rf{js} self-correlator, from which we find the low-frequency topological conductivity according to Eq.~\rf{vs}:
\eq{
\sigma(\omega)=\frac{i}{\omega}\left(-\frac{\chi_s}{(\pi\chi)^2}+\frac{1}{\pi^2\chi}\right)=\frac{\tau}{\pi^2\chi}\,.
}
The response is Drude-like, with a well-defined DC limit and a Drude weight $\propto\chi^{-1}$ (in the $\tau\to\infty$). In the Gilbert damping phenomenology of magnetic dynamics,\cite{gilbertIEEEM04} the spin-pumping rate (i.e., torque $\tau_\alpha$) into the environment is given by
\eq{
-\frac{ds}{dt}\equiv\tau_\alpha=\alpha\p_t\varphi=\alpha\frac{s-s_0}{\chi}\,,
}
which can be obtained by supplementing the Heisenberg-Hamilton dynamics of the planar order parameter with the Rayleigh dissipation function $\mR=\alpha(\p_t\varphi)^2/2$.\cite{tserkovPRB17} We are defining the Gilbert constant $\alpha$ here in units of spin density. This gives $\tau=\chi/\alpha$ for spin-relaxation time, resulting in the low-frequency winding conductivity of
\eq{
\sigma=\frac{1}{\pi^2\alpha}\,.
\label{Gs}
}
A larger damping implies lower winding conductivity.

The corresponding electrical transconductance \rf{G}, $G\propto(\alpha L)^{-1}$, reproduces the spin-superfluid mediated transconductance (drag) derived in Ref.~\onlinecite{takeiPRL14}, when $L\to\infty$. Indeed, in the limit of no axial anisotropy $K$ in Eq.~\rf{Hs} or when it renormalizes down by quantum fluctuations, the out-of-plane spin $s$ is conserved and follows a superfluid-like hydrodynamics (with spin flux $\propto\partial_x\varphi$). This provides a dual description for the long-range signal propagation along the spin chain. In general, however, when the spin is not conserved, the winding hydrodynamics based on Eq.~\rf{CEw} establishes a more universal framework for low-frequency long-distance transport in anisotropic spin chains. This is illustrated in our next and final example.

\subsection{Finite temperatures}
\label{lld}

In the opposite, classical limit of $S\gtrsim1$ and thus $g<2$ in Eq.~\rf{dyg}, the phase $\varphi$ is pinned and the winding dynamics are gapped. In this case, the zero-temperature transport is exponentially suppressed over a long spin chain.\cite{hoffmanCM18} At finite temperatures, however, we generally anticipate a diffusive regime for thermally-activated chiral domain walls,\cite{kimPRB15br} along with the parasitic thermal phase slips.\cite{kimPRB16} If the latter is governed by a larger energy gap, in comparison to the domain-wall energy (which is the case when $K_\perp>K$), we can have a meaningful transport scenario for the conserved winding carried by the Brownian motion of domain walls, along a finite-length spin chain.\cite{kimPRB15br}

In the strong easy-plane limit, when the winding is carried by a classical gas of stable solitonic domain walls (of width $\lambda=\sqrt{A/K}$) with quantized topological charge $\pm1$ and mobility $M$, the corresponding conductivity is simply $\sigma=2n M$, where $n$ is the density of domain walls of a given chirality. The associated diffusion coefficient is given by $D=k_BTM$, according to the Einstein-Smoluchowski relation. Within the Gilbert-damping phenomenology,\cite{gilbertIEEEM04} the mobility of a rigid soliton is given by $M\sim\lambda/\alpha$. Since $\sigma\propto n\propto e^{-\beta E}$,\cite{kimPRB15br} while $\kappa\propto e^{-\beta F}$,\cite{kimPRB16} where $E=4\sqrt{AK}$ is the domain-wall energy and $F=4\sqrt{AK_\perp}\gg E$ is the thermal phase-slip barrier, we can easily satisfy Eq.~\rf{L} at low temperatures. In the limit of $K\to0$ (and/or high temperature, $k_BT\gtrsim E$), the domain walls coalesce and we reproduce the conductivity \rf{Gs}.\cite{kimPRB15br}

\subsection{Noise and quantum relaxometry}
\label{n}

In addition to an electrical measurement of winding transport, as sketched in Fig.~\ref{circ}, it may be possible to investigate the associated topological transport properties, such as winding conductivity, using quantum-impurity (such as nitrogen-vacancy) relaxometry.\cite{duSCI17,*casolaNATRM18} Similarly to the Johnson-Nyquist noise generally associated with charge conductivity, the winding transport is noisy. In particular, the out of the easy-plane spin fluctuations (being canonically conjugate to the planar spin precession, irrespective of the nature of the spin order), should produce a detectable magnetostatic signal.\cite{Note7} We expect it to reflect similar winding transport properties as the electrical setup of Fig.~\ref{circ} (without the issues pertaining to the contacts), in the long-wavelength low-frequency limit of the dynamics. The latter can be controlled by the quantum-impurity positioning and applied magnetic field (e.g., Zeeman splitting of the nitrogen-vacancy spin states), respectively.\cite{flebusPRL18}

\section{Discussion}
\label{d}

In summary, we have constructed a general framework to study winding dynamics in spin chains, from the perspective of a transport phenomenon. Motivated by the mean-field considerations that draw on the notions of spin superflows along the chain and parasitic vorticity flows transverse to it, we developed a fully quantum theory, where both the winding transport and its relaxation by phase slips can be treated systematically by field-theoretical approaches.

We illustrated the general formalism by specializing to the case of antiferromagnetic easy-plane dynamics, whose salient features can be captured by a sine-Gordon model. Two distinct scenarios then arise concerning the winding flows: the spin-superfluid regime, where the parasitic in-plane anisotropy is washed out by quantum (or thermal) fluctuations, and the solitonic regime, where chiral domain walls carry conserved winding density by Brownian motion (at finite temperatures). Both of these limits are addressed within our general Kubo formalism, reproducing and complimenting the pertinent special cases known in the literature. We see that rather generically, in the presence of magnetic damping, the winding flow can exhibit Drude-like dynamic response. This corresponds to an effectively metallic behavior of the conserved winding transport. The key internal-consistency check for these findings concerns the transverse vorticity flow, which reflects phase slips and needs to be smaller than the winding flow along the spin chain.

One of our central motivations for this work is the potential ability to detect the topological transport, either in an electrical device (cf. Fig.~\ref{circ}) or by a nonintrusive quantum-impurity relaxometry (cf. Sec.~\ref{n}). The field-theoretical framework combined with the experimental tangibility should open gates for nonelectrical transport-based investigations of correlated magnetic materials. It is useful to add, furthermore, that a long-range order of any kind is neither assumed nor needed for the emergence of topological hydrodynamics. Our microscopic quantum formulation, which we have explicitly constructed for vorticity and winding flows, furthermore, does not even rely on a local ordering or any semiclassical approximations.

We have largely left out the contact-impedance considerations in our device concept sketched in Fig.~\ref{circ}. This may be justified when there a finite bulk resistivity for the topological flow. In the opposite, clean limit, the transport physics would, however, generally be dominated by the contacts and, at low temperatures, potentially strongly dependent on the many-body effects away from the contacts. These aspects are left for future work. No attempt has been made to classify scenarios of topological hydrodynamics for general quantum magnets in arbitrary dimensions, which also goes beyond our scope here. Quantum skyrmions in two spatial dimensions\cite{ochoaIJMPB19} and hedgehogs in three dimensions\cite{zouCM20} provide other interesting examples, with skyrmions, like winding, obeying only an approximate continuity equation. We thus anticipate rich possibilities for topological hydrodynamics in magnetic materials, with implications for novel probes and device concepts that do not rely on electronic (charge) transport.

\begin{acknowledgments}
We thank Michael Mulligan for insightful discussions. The work was supported by NSF under Grant No. DMR-1742928 (Y.T. and J.Z.). S.K.K. was supported by Brain Pool Plus Program through the National Research Foundation of Korea funded by the Ministry of Science and ICT (Grant No. NRF-2020H1D3A2A03099291) and by the National Research Foundation of Korea funded by the Korea Government via the SRC Center for Quantum Coherence in Condensed Matter (Grant No. NRF-2016R1A5A1008184). S.T. is supported by CUNY Research Foundation Project \#90922-07~10 and PSC-CUNY Research Award Program \#63515-00~51.
\end{acknowledgments}

\end{document}